\documentclass[pra,letterpaper,10pt,twocolumn]{revtex4}
\usepackage{graphicx}
\usepackage[final,dvips]{epsfig}
\usepackage{color}
\usepackage{dcolumn}  
\usepackage{bm}       
\usepackage{array}    
\usepackage{amssymb,amsmath}
\usepackage[colorlinks,citecolor=red]{hyperref}
\emergencystretch=\maxdimen
\include{MyCommand}

\begin{document}

\title{
Effects of an Additional Conduction Band on 
Singlet-Antiferromagnet Competition in the Periodic Anderson Model
}

\author{Wenjian Hu$^1$}
\author{Richard T. Scalettar$^1$}
\author{Edwin W. Huang$^{2,3}$}
\author{Brian Moritz$^{3,4}$}
\affiliation{
$^1$Department of Physics, University of California Davis, Davis, CA 95616, USA \\
$^2$Department of Physics, Stanford University, Stanford, CA 94305, USA \\
$^3$Stanford Institute for Materials and Energy Sciences, SLAC National
Accelerator Laboratory and Stanford University, Menlo Park, CA 94025, USA \\
$^4$Department of Physics and Astrophysics, University of North Dakota,
Grand Forks, ND 58202, USA
}

\begin{abstract}
The competition between antiferromagnetic
(AF) order and singlet formation is a central phenomenon of the
Kondo and Periodic Anderson Hamiltonians, and of the heavy fermion
materials they describe.
In this paper, we explore the effects of an additional conduction band
on magnetism in these models, and, specifically, on changes in
the AF-singlet quantum critical point (QCP) and the one particle and
spin spectral functions.
To understand the magnetic phase transition qualitatively, we first
carry out a self-consistent mean field theory (MFT).  The basic
conclusion is that, at half-filling, the coupling to the additional band
stabilizes the AF phase to
larger $f$ $d$ hybridization $V$ in the PAM.  We also explore the possibility
of competing ferromagnetic phases when this conduction band is
doped away from half-filling. 
We next employ Quantum Monte Carlo (QMC)
which, in combination with finite size scaling, allows us to
evaluate the position of the QCP using an exact treatment of the interactions.
This approach confirms the stabilization of AF order, which occurs
through an enhancement of the Ruderman-Kittel-Kasuya-Yosida (RKKY)
interaction.  QMC results for
the spectral function $A(\textbf{q},\omega)$ and dynamic spin
structure factor $\chi(\textbf{q},\omega)$ yield additional
insight into the AF-singlet competition and the low temperature phases.
\end{abstract}

\maketitle

\section{Introduction}
The periodic Anderson Model (PAM) describes the hybridization between
mobile ($d$ band) free electrons in a metal with  strongly correlated
$f$ electrons.  The PAM has been extensively studied since its first
introduction\cite{Anderson61}, and can successfully account for a variety
of remarkable $f$-electron (rare-earth and actinide) phenomena including
heavy-fermion
physics\cite{stewart84,lee86,Georges96,Vidhyadhiraja04,Sen16}, valence
fluctuations\cite{hewson93,Shinzaki16}, volume collapse
transitions\cite{allen82,allen92,lavagna83,gunnarsson83,mcmahan98,lipp08,bradley12,lanata15}
and unconventional superconductivity\cite{pixley15}. 

At low temperatures, as the hybridization strength is varied in the PAM,
there is a competition between the RKKY
interaction\cite{Ruderman54,Kasuya56,Yosida57}, which favors
magnetically ordered $f$ band local moments, and the Kondo
effect\cite{Kondo69, Kouwenhoven01}, which screens the local moments and
induces singlet states.  Kondo screening can also occur at the interface
between metallic and strongly correlated
materials\cite{mannhart05,mannhart10}, a situation which has given rise
to additional theoretical and numerical investigation of the PAM and its
geometrical variants\cite{millis05,freericks06,Euverte12}.  Here the
metallic band is viewed as arising from material on one side of an
interface, and the correlated band describes the other side of the
interface, as opposed to originating from strongly and weakly correlated
orbitals of atoms in a single, homogeneous material.

A natural generalization of models which couple a single conduction
band to localized, magnetic, orbitals is to consider similar physics
when several conduction bands are present.  The new qualitative
physics to be explored is how the third band, and the resulting
imbalance between the numbers of conduction and loacalized electrons,
alters the strong correlation phenomena of the two band PAM: 
RKKY-induced AF
order at weak $V$, the nature of the Kondo gap at strong $V$, and,
finally, the position of the AF-singlet transition between these limits.


Besides these interesting fundamental questions, such a model is also
worthy of investigation as a first step towards experiments\cite{shishido10} on
$f$-electron superlattices like CeIn$_3$(n)/LaIn$_3$(m).  In these
systems, by varying the
thicknesses $n,m$ of the different materials, the effective
dimensionality can be tuned, and hence the $2D$-$3D$ crossover of Kondo
physics and AF order.  


Recent theoretical investigations of the effect of immersing a
Kondo insulator, or a superlattice thereof, in a 3D metal has been
undertaken by Peters {\it et al.}\cite{peters13}.  The focus there was on
the evolution of the density of states $\rho(\omega)$ and, especially,
features like the Fermi level hybridization gap of the Kondo sheet.  A
key conclusion is that the Kondo gap is modified to a pseudogap, with
quadratically vanishing $\rho(\omega)$ from coupling to the metallic
layer, and that the 3D $\rho(\omega)$ of the metallic layer adjacent to
the Kondo layer develops 2D features.  Changes to $\rho(\omega=0)$
in the singlet phase will be a key feature of our results here.

A bilayer heavy fermion system
comprising a Kondo insulator (KI), represented by a symmetric PAM,
coupled to a simple metal (M) has been proposed and studied
employing the framework of DMFT\cite{Sen16} at half-filling.  
The main goal of the work
was to determine the ground state phase diagram from a Kondo screened
Fermi liquid to a Mott insulating phase as a function of interaction
strength and interlayer coupling.  More generally, the possibility of
the coexistence of spectral functions with distinct behaviors near the
Fermi surface, despite the presence of interband hybridization, is the
topic of studies of orbitally selective
transitions\cite{liebsch04,arita05}.

While we focus here on the influence of an additional metallic band on
the properties of the PAM, similar extensions to include electron-phonon
coupling\cite{Zhang13}, dilution\cite{Sen15}, and $f$-electron
hybridization\cite{Euverte13} have similarly explored the ways in which
AF-singlet competition can be influenced by the inclusion of further
energy scales and degrees of freedom in the Hamiltonian.

In this paper, we employ the determinant quantum Monte Carlo (DQMC)
method\cite{Blankenbecler81, Loh92}, which provides an
approximation-free solution to strong correlations, to study the
magnetic structure of the bilayer KI-M system. By finite size
scaling, we reliably extract the AF order
parameter as a function of the KI hybridization strength $V$, and then
build up the magnetic phase diagram in the $V-t'$ plane for a
representative potential $U_f=4$.  To understand more
precisely the role of nonzero $t'$, we begin with a redetermination of
the quantum critical point (QCP) of the AF-singlet phase transition of
the half-filled PAM, the $t'=0$ limit, with higher accuracy than in
previous literature\cite{Vekic95}. The DQMC work is mainly focused on the 
particle-hole symmetric (half-filled) limit where there is no
sign problem in the simulation. We also implemented a mean-field theory (MFT) 
calculation bothe at and away from half-filling as a supplement to DQMC. 
Our work is distinguished from
previous work\cite{peters13,Sen16} by its consideration of a PAM rather
than a coupling to local (Kondo) spins, and its treatment of intersite
magnetic correlations which are suppressed in the paramagnetic DMFT used in
earlier work.  

\section{Model and Methods}
\begin{figure}[!h]
\includegraphics[width=0.8\columnwidth]{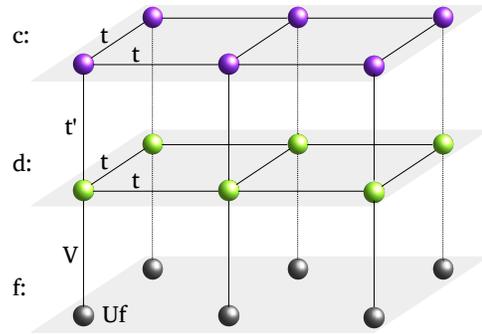} 
\caption{
Three band system with $f$ and $d$ bands comprising the KI (PAM) layer
coupling to an additional $c$ (metallic) layer.  These KI and metal are
coupled through an interlayer hopping parameter $t'$.  Our primary
tuning parameter will be the hybridization $V$ within the KI.
\label{fig:geometry}
}
\end{figure}

We consider the bilayer KI-M Hamiltonian on a square lattice,
\begin{align}
H &= -t \sum_{<\textbf{i},\textbf{j}>\sigma}(c^{\dagger}_{\textbf{i}\sigma}c^{\phantom{\dagger}}_{\textbf{j}\sigma}+
c^{\dagger}_{\textbf{j}\sigma}c^{\phantom{\dagger}}_{\textbf{i}\sigma})
+\epsilon^c\sum_{\textbf{i}\sigma}n^c_{\textbf{i}\sigma} \nonumber \\
& -t\sum_{<\textbf{i},\textbf{j}>\sigma}(d^{\dagger}_{\textbf{i}\sigma}d^{\phantom{\dagger}}_{\textbf{j}\sigma}
+d^{\dagger}_{\textbf{j}\sigma}
d^{\phantom{\dagger}}_{\textbf{i}\sigma})+\epsilon^d\sum_{\textbf{i}\sigma}
n^d_{\textbf{i}\sigma} \nonumber \\
& +U_f\sum_{\textbf{i}}(n^f_{\textbf{i}\uparrow}-\frac{1}{2})(n^f_{\textbf{i}\downarrow}-\frac{1}{2}) +\epsilon^f\sum_{\textbf{i}\sigma}n^f_{\textbf{i}\sigma} \nonumber \\
& -t'\sum_{\textbf{i}\sigma}(c^{\dagger}_{\textbf{i}\sigma}d^{\phantom{\dagger}}_{\textbf{i}\sigma}
+d^{\dagger}_{\textbf{i}\sigma}
c^{\phantom{\dagger}}_{\textbf{i}\sigma})-V\sum_{\textbf{i}\sigma}(d^{\dagger}_{\textbf{i}\sigma}f^{\phantom{\dagger}}_{\textbf{i}\sigma}
+f^{\dagger}_{\textbf{i}\sigma}d^{\phantom{\dagger}}_{\textbf{i}\sigma})
\label{eq:ham}
\end{align}
$t$ is the intralayer 
hopping parameter, which, for simplicity, we chose to be the
same in the uncorrelated $c$ and $d$ bands. $t'$ is the interlayer
hopping parameter between the $c$, $d$ bands. $V$ is the hybridization
strength between the $d$, $f$ bands. $U_f$ is the Coulomb repulsion in
the $f$ band.  Finally, $\epsilon^\alpha$ are the orbital energies of the
$\alpha=c, d,$ and $f$ bands, and $n^\alpha_{\textbf{i}\sigma}\equiv
\alpha^{\dagger}_{\textbf{i}\sigma}
\alpha^{\phantom{\dagger}}_{\textbf{i}\sigma}$ are the density
operators. The model is shown pictorially in Fig.~\ref{fig:geometry},
where $f$ and $d$ bands belong to the KI layer, and $c$ band belongs to
the metal layer. 
Within the KI formed by the $d$ and $f$ bands, $V$ controls the
competition between antiferromagnetic (AF) and singlet phases.  

In this work, we set $t=1$ as our energy scale and mainly consider the
particle-hole symmetric limit where $\epsilon^\alpha=0$, so each of the
three bands is individually half-filled. We also implement the MFT
calculation away from half-filling. 

At half-filling, the
Hamiltonian can be solved exactly in the noninteracting limit ($U_f=0$).
Unlike the PAM in which $V$ opens a gap at half-filling and which hence
is a band insulator there, the KI-M system is metallic at half-filling.
This follows from the fact that the Hamiltonian has an odd number of
bands (three): the Fermi level lies in the middle of the
central band.

This metallic character at half-filling can be made more precise
by going to momentum space 
$\alpha^{\dagger}_{\textbf{k}\sigma}=
( 1 \ \sqrt{N}) \,
\sum_\textbf{l}
e^{i\textbf{kl}}\alpha^{\dagger}_{\textbf{l}\sigma}$ for each of the three
bands $\alpha=c,d,f$.
\begin{align}
 \begin{small}
  H=\sum_{\textbf{k}\sigma}
  \left[ {\begin{array}{ccc}
   c^{\dagger}_{\textbf{k}\sigma} & d^{\dagger}_{\textbf{k}\sigma} & f^{\dagger}_{\textbf{k}\sigma} \\
  \end{array} } \right]
  \left[ {\begin{array}{ccc}
   \epsilon_{\textbf{k}} & -t' & 0\\
   -t' & \epsilon_{\textbf{k}} & -V\\
   0 & -V & 0\\
  \end{array} } \right]
  \left[ {\begin{array}{c}
   c_{\textbf{k}\sigma} \\ d_{\textbf{k}\sigma} \\ f_{\textbf{k}\sigma} \\
  \end{array} } \right]
 \end{small}
\label{eq:ham_Uf=0}
\end{align}
Here $\epsilon_{\textbf{k}}=-2t(\cos(k_x)+\cos(k_y))$. 
Diagonalizing the Hamiltonian Eq.~\ref{eq:ham_Uf=0} yields the three
energy bands.  In general, these bands cross and, at $\epsilon^\alpha=0$, are
all partially filled.  However, in certain limits, e.g.~$V=0$ and
$t'>4t$, band gaps are present.  Even so, in these
situations the central band is half-filled and the system remains
metallic.

The focus of our work will be the implications of the additional
metallic ($c$) band on the competition between the RKKY
interaction-induced AF  and the Kondo regime of screened singlets, both
central to the behavior of heavy-fermion
materials\cite{stewart84,lee86,Georges96} and 
captured in the solution of the PAM ($t'=0$).  A
natural expectation is that, with the increase of $t'$, the RKKY
interaction between the local $f$ moments is enhanced, due to the
additional conduction band channels, while the Kondo energy scale, set
by $V$ and $U_f$, remains roughly fixed.  This should lead to an overall
movement of the quantum critical point to larger values of $V$.

To understand the precise effect of $t'$ on the AF-singlet transition, we first
carried out a self-consistent mean field theory (MFT).  We then turned
to a more exact, DQMC solution.

\section{Results:  Mean Field Theory}

Together with Kondo phases,
ferromagnetic, antiferroferromagnetic and mixed order are all
possible in the PAM and related Hamiltonians\cite{bernhard15,eder16}.
In the MFT treatment presented here, we thus consider three possible
phases, the AF phase, the ferromagnetic (F) phase and the singlet phase.
The AF MFT {\it ansatz} is
\begin{align}
\langle n^f_{\textbf{l}\sigma}\rangle=\frac{n_f}{2}+
\frac{\sigma m_f (-1)^{\textbf{l}}}{2}
\label{eq:MFT_AFM}
\end{align}
where $\sigma$ is spin up ($\uparrow$) or spin down ($\downarrow$) and
$m_f$ is the $f$ band AF order parameter. While the F MFT {\it ansatz} is
\begin{align}
\langle n^f_{\textbf{l}\sigma}\rangle=\frac{n_f}{2}+
\frac{\sigma m_f}{2}
\label{eq:MFT_FM}
\end{align}  
where $\sigma$ follows the same definition and $m_f$ is the $f$ band
ferromagnetic order parameter. In order to fix the particle densities
$n_c$, $n_d$ and $n_f$, the terms $N n_c \epsilon^c$, $N n_d \epsilon^d$
and $N n_f \epsilon^f$ must be subtracted from the original Hamiltonian,
Eq.~\ref{eq:ham} reported in Ref.~\cite{Costaa}.

The AF MF decoupling of the interaction then gives a quadratic
Hamiltonian in which momenta ${\bf k}$ and ${\bf k}$ - ${\bf Q}$ (where
${\bf Q}=(\pi,\pi)$) are coupled, resulting in
\begin{align}
  H_{AF}&=\frac{1}{2}\sum_{\textbf{k}\sigma}v^{\dagger}_{\textbf{k}\sigma}
  M^{AF}_{\textbf{k}\sigma}
  v^{\phantom{\dagger}}_{\textbf{k}\sigma}+\frac{NU_f(m^2_f + n^2_f - 1)}{4}  \nonumber\\
  &-N n_c \epsilon^c - N n_d \epsilon^d - N n_f \epsilon^f
\label{eq:MFT_H_AFM}
\end{align}
where $v^{\dagger}_{\textbf{k}\sigma}=\left[ {\begin{array}{cccccc}
   c^{\dagger}_{\textbf{k}\sigma} & d^{\dagger}_{\textbf{k}\sigma} & f^{\dagger}_{\textbf{k}\sigma} &
   c^{\dagger}_{\textbf{k}-\textbf{Q},\sigma} & d^{\dagger}_{\textbf{k}-\textbf{Q},\sigma} & f^{\dagger}_{\textbf{k}-\textbf{Q},\sigma} \\
  \end{array} } \right]$
, 
\begin{align}
  M^{AF}_{\textbf{k}\sigma}=\left[ {\begin{array}{cccccc}
   \eta^c_{\textbf{k}} & -t' & 0 & 0 & 0 & 0\\
   -t' & \eta^d_{\textbf{k}} & -V  & 0 & 0 & 0\\
   0 & -V & \eta^f_{\textbf{k}}  & 0 & 0 & -\frac{\sigma m_f U_f}{2} \\
   0 & 0 & 0 & \eta^c_{\textbf{k}-\textbf{Q}} & -t' & 0\\
   0 & 0 & 0 & -t' & \eta^d_{\textbf{k}-\textbf{Q}} & -V \\
   0 & 0 & -\frac{\sigma m_fU_f}{2} & 0 & -V & \eta^f_{\textbf{k}-\textbf{Q}} \\
  \end{array} } \right] .
\nonumber
\end{align}
In $M^{AF}_{\textbf{k}\sigma}$, $\eta^c_{\textbf{k}}$,
$\eta^d_{\textbf{k}}$ and $\eta^f_{\textbf{k}}$ stand for
$\epsilon_{\textbf{k}}+\epsilon^c$, $\epsilon_{\textbf{k}}+\epsilon^d$
and $(\frac{n_f}{2}-\frac{1}{2})U_f+\epsilon^f$ respectively.

On the other hand, the F MF decoupling leads to
\begin{align}
  H_{F}&=\sum_{\textbf{k}\sigma}u^{\dagger}_{\textbf{k}\sigma}
  M^{F}_{\textbf{k}\sigma}
  u^{\phantom{\dagger}}_{\textbf{k}\sigma}+\frac{NU_f(m^2_f + n^2_f - 1)}{4} 
  \nonumber\\
  &-N n_c \epsilon^c - N n_d \epsilon^d - N n_f \epsilon^f
\label{eq:MFT_H_FM}
\end{align}
where $u^{\dagger}_{\textbf{k}\sigma}$ stands for $\left[ {\begin{array}{ccc}
   c^{\dagger}_{\textbf{k}\sigma} & d^{\dagger}_{\textbf{k}\sigma} & f^{\dagger}_{\textbf{k}\sigma} \\
  \end{array} } \right]$ and the matrix is
\begin{align}
  M^{F}_{\textbf{k}\sigma}=\left[ {\begin{array}{ccc}
   \epsilon_{\textbf{k}}+\epsilon^c & -t' & 0 \\
   -t' & \epsilon_{\textbf{k}}+\epsilon^d & -V \\
   0 & -V & (\frac{n_f}{2} - \frac{1}{2} - \frac{\sigma m_f}{2})U_f+\epsilon^f \\
  \end{array} } \right].
\nonumber
\end{align}

Hence, $m_f$, $\epsilon^c$, $\epsilon^d$, $\epsilon^f$ and the
corresponding MFT phase boundary,
are computed self-consistently by minimizing the total ground state energy $E$
\begin{align}
\left\langle \frac{\partial E}{\partial m_f}\right\rangle 
=\left\langle \frac{\partial E}{\partial \epsilon^c}\right\rangle 
=\left\langle \frac{\partial E}{\partial \epsilon^d}\right\rangle 
=\left\langle \frac{\partial E}{\partial \epsilon^f}\right\rangle 
= 0.
\end{align}
\begin{figure}[!h]
\includegraphics[width=0.98\columnwidth]{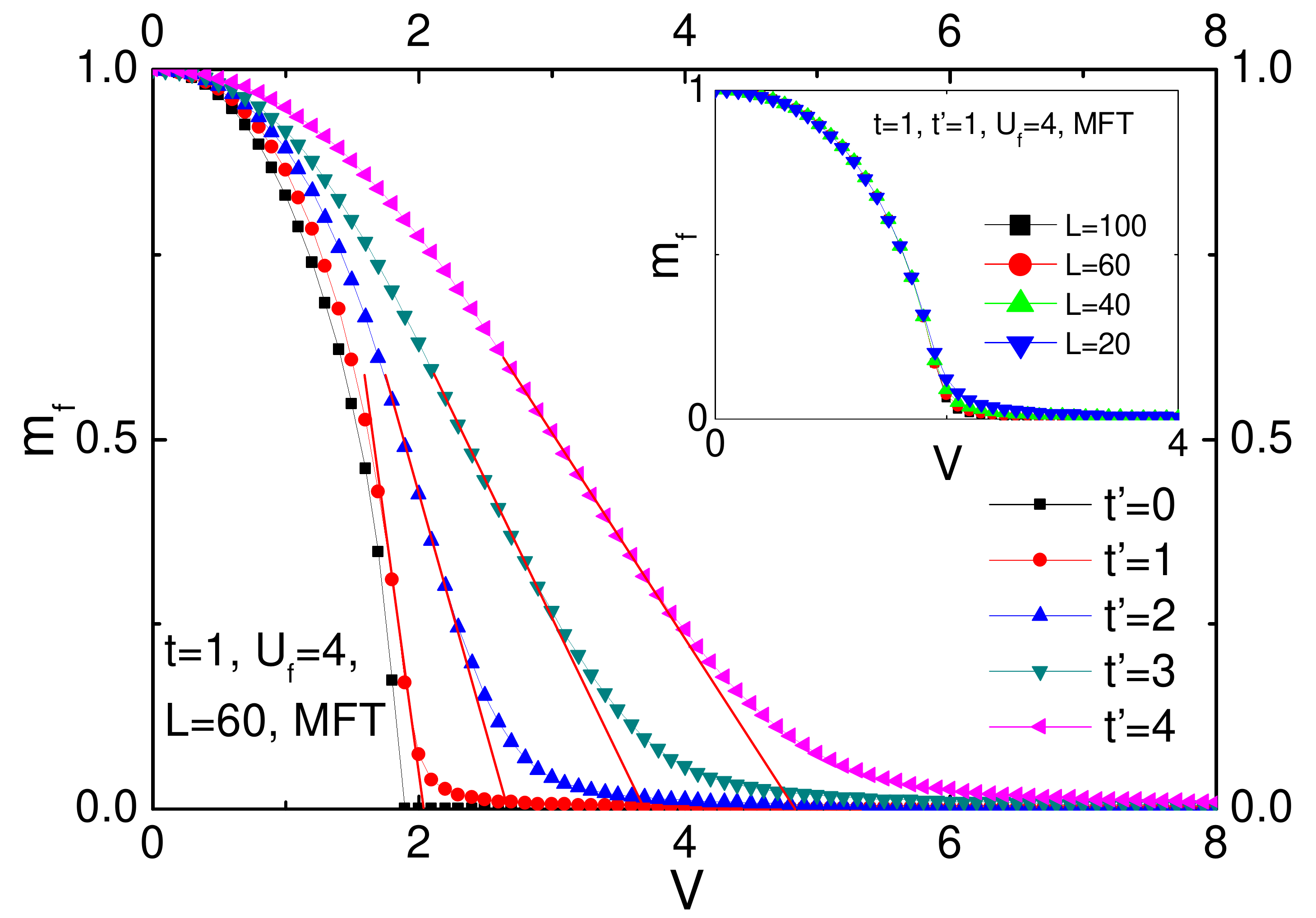} 
\caption{
MFT results for the AF order parameter $m_f$ as a
function of $V$ for several KI-M hybridizations, $t'=0,1,2,3,4$. 
Each band is fixed at half-filling. 
The on-site interaction in the KI, $U_f=4$, is fixed.  At $t'=0$, where the
system is insulating in the non-interacting limit, the order parameter
rises from zero steeply, i.e.~with the expected exponent $\beta=1/2$ for
$V< V_c$.  For $t' \neq 0$, $m_f$ exhibits a more gradual behavior.
This smoother cross-over is a consequence of the diverging density of
states at the Fermi surface, and reflects the persistence of AF order at
large $V$.  A `cross-over value' $V_c$ is identified as described in
the text.  Calculations shown are for lattice sizes of linear extent
$L=60$, and verified to have converged. (See inset.) 
\label{fig:MFT_smooth}
}
\end{figure}

We first explore the AF MFT {\it ansatz} at half-filling. 
Results for $m_f$ as a function of $V$ for different
couplings $t'$ of the KI to the metal are shown in 
Fig.~\ref{fig:MFT_smooth}.
A sharp QCP is evident at $t'=0$ whose location agrees with previous
work\cite{Vekic95}.  The evolution of $m_f(V)$ is smoother
for $t' \neq 0$.  This difference is associated with the  fact that
at $t'=0$ the KI is a band insulator in the noninteracting limit,
whereas, as discussed in the previous section, the KI-M
model is metallic.  In fact, the density of states $N(E)$ has
a van-Hove singularity at the half-filled Fermi
surface $E_{\bf k}=0$, for all $V$ with $t' \neq 0$, as also
occurs in the square lattice half-filled Hubbard model. 
In an expansion of the free energy, 
$F(m_f)$ picks up
a $|m_f|$ contribution from these $E_{\bf k}=0$ modes,
which persists in the thermodynamic limit owing to 
the divergence of $N(E=0$).
This effect
pushes $V_c$ out to $V=\infty$ in mean field theory:
AF order persists for all hybridization strengths.

Nevertheless, a cross-over $V_c$ is still evident in 
Fig.~\ref{fig:MFT_smooth}, especially for modest $t'$.
We assign a quantitive value by choosing the point of 
$m_f(V)$ of largest slope, and extrapolating linearly to
$m_f=0$ as shown.  These cross-over values for $V_c$ will be compared
with the critical hybridization obtained by DQMC in the 
following section.

\begin{figure}[!h]
\includegraphics[width=0.98\columnwidth]{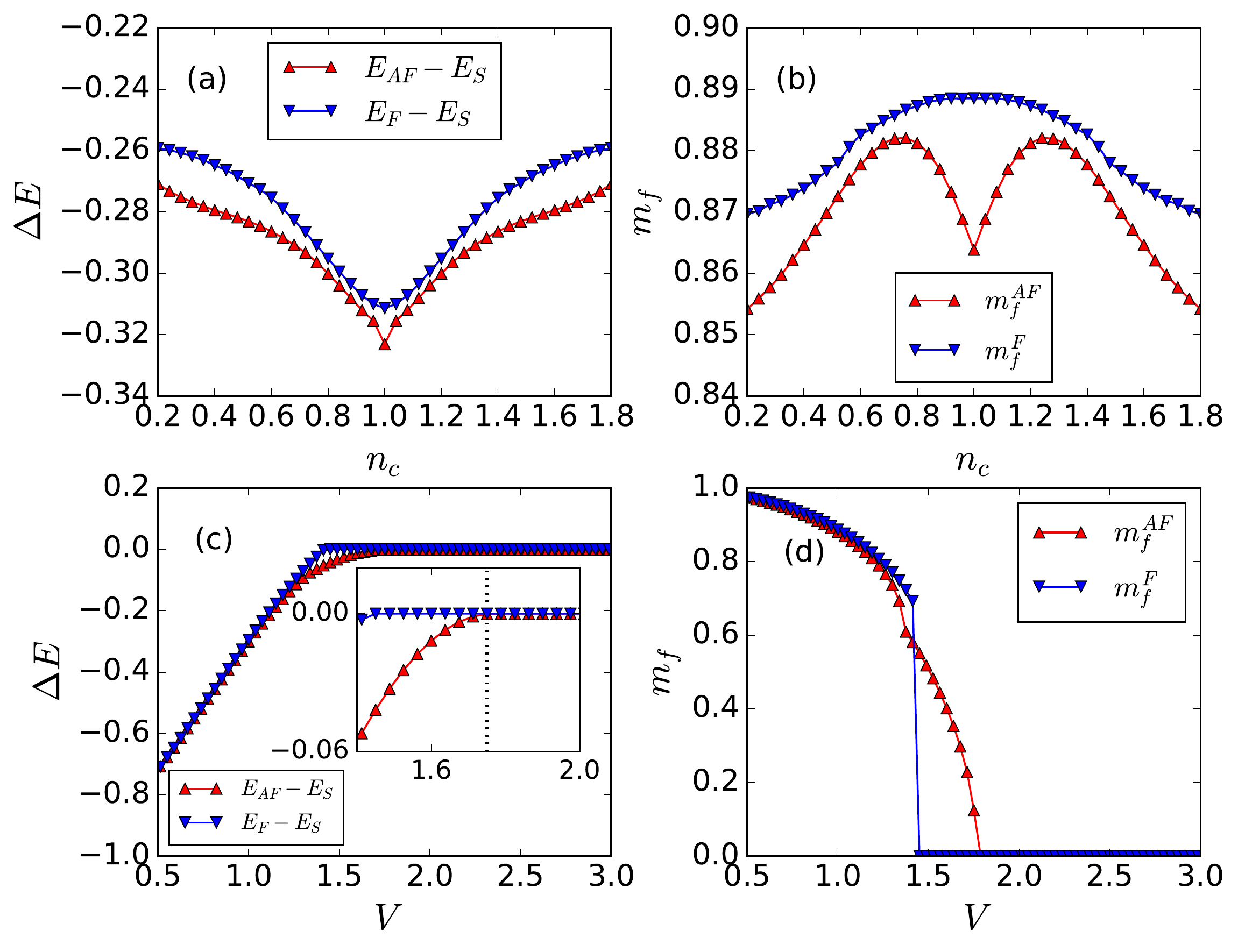} 
\caption{
MFT results away from half-filling. (a) $E_{AF}-E_{S}$ and $E_{F} -
E_{S}$ as a function of density $n_c$. (b) $m^{AF}_f$ and $m^{F}_f$ as a
function of density $n_c$. Panels (a) (b) are fixed at $n_d=1$, $n_f=1$,
$t'=1$, $V=1$, $U_f=4$ and $L=200$. (c) $E_{AF}-E_{S}$ and $E_{F} -
E_{S}$ as a function of hybridization strength $V$. The dashed line in the inset
indicates the critical point $V_c$. (d) $m^{AF}_f$ and
$m^{F}_f$ as a function of $V$. Panels (c) (d) are fixed at $n_c=0.8$,
$n_d=1$, $n_f=1$, $t'=1$, $U_f=4$ and $L=200$.
\label{fig:figure_MFT}
}
\end{figure}

To determine the ground state phase away from half-filling, we compare
results from both the AF MFT {\it ansatz} and the F MFT {\it ansatz}, as
shown in Fig.~\ref{fig:figure_MFT}. We denote the corresponding
ground state energies
$E_{AF}$ ($E_{F}$) and, for the singlet (paramagnetic) phase,
$E_{S}$. In
Fig.~\ref{fig:figure_MFT} (a), $E_{AF}-E_{S}$ and $E_{F}-E_{S}$ are
shown as a function of the density $n_c$ with fixed $n_d=1$, $n_f=1$,
that is, by doping the additional conduction band.  We have chosen
$t'=1$, $V=1$, $U_f=4$ and $L=200$.  For these
parameters, $E_{AF}$ is always lowest for all dencities $n_c$.
In Fig.~\ref{fig:figure_MFT}(b), the optimal $m_f$ from
the AF MFT {\it ansatz} (denoted as $m^{AF}_f$) and the optimal $m_f$
from the F MFT {\it ansatz} (denoted as $m^{F}_f$) are shown as a
function of density $n_c$.
Since the ground state phase is AF, red
triangular data characterizes the behaviour of the ground state order
parameter. While $n_c$ is varied greatly from $0.2$ to $1.8$, the
magnitude of the order parameter stays in a small range from $0.85$ to
$0.89$, showing that the $c$ band has limited effects on the $f$ band
magnetic structure. 

In Fig.~\ref{fig:figure_MFT}(c), $E_{AF}-E_{S}$ and $E_{F}-E_{S}$ are
shown as a function of the hybridization strength $V$ with fixed
$n_c=0.8$, $n_d=1$, $n_f=1$, $t'=1$, $U_f=4$ and $L=200$. Below the
critical point $V_c\approx 1.75$, $E_{AF}$ is lower than $E_{F}$ and
$E_{S}$. Above the critical point,
the AF order gives way to the singlet phase in a second order
phase transition. In Fig.~\ref{fig:figure_MFT}(d), the explicit
behaviours of $m^{AF}_f$ and $m^{F}_f$ with respect to $V$ are
presented, in agreement with results of
Fig.~\ref{fig:figure_MFT}(c). Red triangular data points characterize
the behaviour of the ground state order parameter. Notably, by moving
away from half-filling, the smooth transition observed at the
half-filling limit returns to the conventional MFT transition behaviour with
order parameter exponent $\beta = 1/2$.

\section{Results:  Determinant Quantum Monte Carlo}

In contrast to MFT, DQMC provides an exact treatment of the 
interactions in the KI-M Hamiltonian.  This is accomplished through
the construction of a path integral expression for the partition
function and the introduction of an auxiliary field 
to decouple the exponential of the quartic interaction term
into a quadratic form\cite{Blankenbecler81}.  The fermion
trace can be done exactly, and the auxiliary field is then sampled to
produce measurements of one and two particle correlation functions.

The DQMC method works on lattices of finite spatial extent,
necessitating an extrapolation to the thermodynamic limit as described
below.  A ``Trotter error" is also introduced in the separation
of the kinetic and potential energy pieces of the Hamiltonian. 
We work with an imaginary time discretization small enough such that
the Trotter error is negligible, i.e.~it is
less than our statistical sampling errors.
All the following DQMC results are presented at the particle-hole symmetric
limit.

To explore the magnetic behaviour, we first study the $f$ band
real space equal time spin-spin correlation function,
\begin{align}
  C^f(\textbf{r})=\langle S^f_{\textbf{i}+\textbf{r}, z}\,
S^f_{\textbf{i},z} \rangle=\langle(n^f_{\textbf{i}+\textbf{r}\uparrow}
-n^f_{\textbf{i}+\textbf{r}\downarrow})
(n^f_{\textbf{i}\uparrow}-n^f_{\textbf{i}\downarrow})\rangle
\label{eq:spsp}
\end{align}
$C^f({\textbf r})$ measures the correlation between the $z$ component of a spin on site
\textbf{i} with that on a site a distance \textbf{r} away. 
Although the definition in Eq.~\ref{eq:spsp} only
involves the $z$ component, we average
all three components (which are equal by rotational
symmetry).

In addition to the spatial decay of the $f$ band spin 
correlation function of Eq.~\ref{eq:spsp}, we also
study the Kondo singlet correlation function\cite{Huscroft99}, 
defined as:
\begin{align}
  C^{fd}=\langle \vec{S}^f_{\textbf{i}}\cdot \vec{S}^d_{\textbf{i}} \rangle
\label{eq:spsp_cfd}
\end{align}
where $\vec{S}^f_{\textbf{i}} = [f^{\dagger}_{\textbf{i} \uparrow} \;
f^{\dagger}_{\textbf{i} \downarrow}] \, {\vec \sigma} \, \left[ \begin{array}{c}
f^{\phantom{\dagger}}_{\textbf{i} \uparrow} \\
f^{\phantom{\dagger}}_{\textbf{i} \downarrow} \end{array}  \right]$ and
$\vec{S}^d_{\textbf{i}} = [d^{\dagger}_{\textbf{i} \uparrow} \;
d^{\dagger}_{\textbf{i} \downarrow}] \, {\vec \sigma} \, \left[ \begin{array}{c}
d^{\phantom{\dagger}}_{\textbf{i} \uparrow} \\
d^{\phantom{\dagger}}_{\textbf{i} \downarrow} \end{array}  \right]$ and
\textbf{$\vec \sigma$} are the Pauli matrices.

\begin{figure}[!h]
\includegraphics[width=0.98\columnwidth]{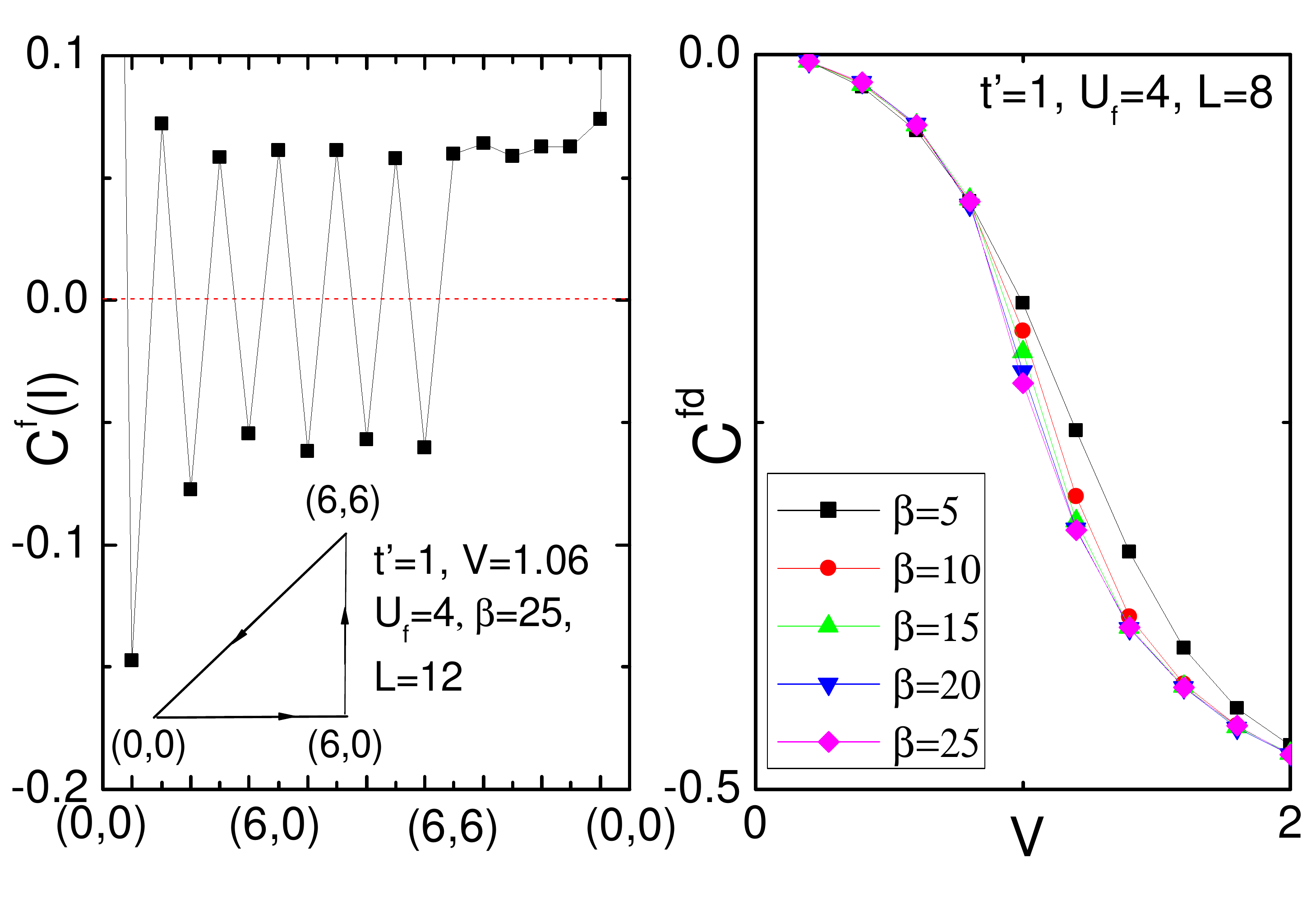} 
\caption{
\underline{Left}: The equal time spin-spin correlation function as a function of distance
$\textbf{r}$, on a $12\times 12$ lattice, with fixed $t'=1$, $V=1.06$,
$U_f=4$, $\beta=25$. The horizontal axis follows the direction of the
triangular path on the lattice with the AF correlations clearly
present. Long range order is clearly evident, even though
$V$ exceeds the critical value $V_c$ for the AF-singlet transition of
the PAM:  $t'$ nonzero has stabilized AF.
\underline{Right}: The singlet correlation function $C^{fd}$
increases in magnitude with
hybridization $V$.  Since $C^{fd}$ measures local correlations,
its reaches an asymptotic low $T$ value at relatively small
$\beta \approx 10$.
\label{fig:KIM_spsp}
}
\end{figure}

At a KI-M coupling $t'=1$, 
the DQMC result for the
$f$ band spin-spin correlation function $C^f(\textbf{r})$ shown
in Fig.~\ref{fig:KIM_spsp} reveals non-zero (long range) 
AF correlations at hybridization strength $V=1.06$. 
This value is well above the pure KI ($t'=0$) critical point,
indicating that AF order is stabilized by $t'$.
The right panel of Fig.~\ref{fig:KIM_spsp}
shows the variation of the singlet correlation function $C^{fd}$ with
hybridization strength. As $V$ increases,
the system switches from a small $C^{fd}$ regime
where singlet correlations are absent (the AF phase dominates) to a
large $C^{fd}$ regime where Kondo singlets are well formed (and AF
correlations are absent).
As has been previously noted\cite{Vekic95}, 
the position of the most rapid increase in magnitude
of $C^{fd}$ gives an approximate location to the AF-singlet QCP.

In addition to the manner in which the spin correlation function decays with
spatial separation, the 
imaginary time evolution also offers a window into the AF-singlet
transition.  Specifically, the $f$ band dynamic local moment,
\begin{align}
  \langle m^2\rangle_{\rm dyn}
&=\frac{1}{\beta}\int^{\beta}_0 d\tau 
\, C^f({\bf r}=0,\tau) 
\nonumber \\
&=\frac{1}{\beta}\int^{\beta}_0 d\tau 
\, \langle S^f_{\textbf{i}z}(\tau)S^f_{\textbf{i}z}(0) \rangle
\label{eq:dy_moment}
\end{align}
is the integral of the spatially local, unequal time
spin correlation function 
$C^f({\bf r}=0,\tau)$.
Here $S^f_{\textbf{i}z}(\tau)=e^{H\tau} S^f_{\textbf{i}z} e^{-H\tau}$.
As with our previous equal time 
$C^f({\bf r})$, we average this correlation function over
all spin directions to improve statistics.
In a situation where the spin operator commutes with the Hamiltonian,
e.g.~at $V=0$ where one has isolated moments, the instantaneous,
$C^f({\bf r}=0,\tau=0)$,
and dynamic moments 
$\langle m^2 \rangle_{\rm dyn}$
are equal.  Quantum fluctuations from the hybridization
$V$ cause the spin correlation to decay in imaginary time, 
reducing the dynamic moment.

\begin{figure}[!h]
\includegraphics[width=0.98\columnwidth]{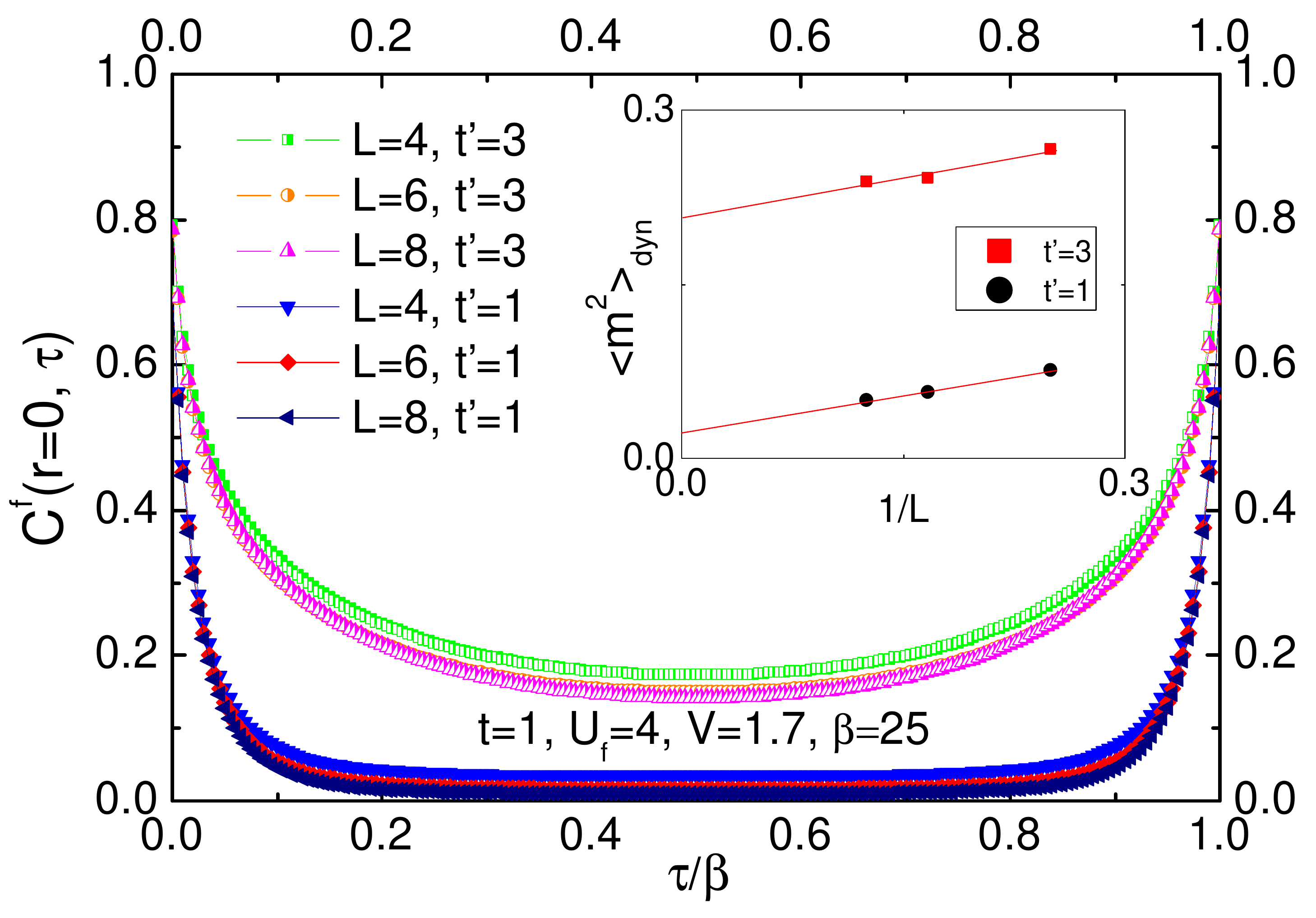} 
\caption{
The spin correlation function as a function of 
imaginary time separation. Here $U_f=4$ and $V=1.70$ are fixed. 
The unequal
time correlation function rapidly decays to zero for $t'=1$, while it
decays smoothly to a non-zero value for $t'=3$, suggesting the
values correspond to two distinct
magnetic phases.  Finite size effects are verified to be
small by comparing data for $L=4,6,8$.  This
lack of dependence on $L$ is associated with the 
fact that the quantity being measured is local in space.
\underline{Inset}: The $f$ band dynamic
local moment as a function of $1/L$. For $t'=1$ the extrapolated local
moment is $\langle m^2 \rangle_{\rm dyn} \sim 0.024$, corresponding to a screened Kondo singlet phase,
while for $t'=3$ it is almost an order of magnitude greater,
$\langle m^2 \rangle_{\rm dyn} \sim 0.21$, corresponding to an ordered magnetic
phase. The extrapolation was performed using a linear least-squares fit.
\label{fig:dy_moment}
}
\end{figure}

As seen in Fig.~\ref{fig:dy_moment} there are two quite
different behaviors of $C^f({\bf r}=0,\tau)$ when $V$ is non-zero.
$C^f({\bf r}=0,\tau)$ decays to zero rapidly at $\tau/\beta = 0.5$ for
$t'=1$, while it decays smoothly to a non-zero value at $\tau/\beta =
0.5$ for $t'=3$. Integrating $C^f({\bf r}=0,\tau)$ yields the dynamic
local moments shown in the inset to Fig.~\ref{fig:dy_moment}. Increasing
$t'$ from $1$ to $3$, induces a very large change in $\langle m^2
\rangle_{\rm dyn}$, which implies the shifting of the system from a
Kondo singlet to an AF phase.

The rapid decay of 
$C^f({\bf r}=0,\tau)$ with $\tau$ is associated with the
presence of a singlet gap.  An alternate way of interpreting the data
of Fig.~\ref{fig:dy_moment} is that by enhancing the AF tendency,
increasing $t'$ causes the vanishing to the singlet gap and 
a large increase in $\langle m^2 \rangle_{\rm dyn}$.

The $f$-band structure factor $S^f(\textbf{k})$ is the Fourier transform
of the $f$-band equal time spin-spin correlation function
$C^f(\textbf{r})$, and is defined as:
\begin{align}
  S^f(\textbf{k})=\sum_{\textbf{r}}e^{i\textbf{k}\textbf{r}}C^f(\textbf{r})
\label{eq:structure}
\end{align}
We present results for $\textbf{k}=\textbf{Q}$, the AF structure factor,
since this is the dominant ordering wave vector at half-filling.

\begin{figure}[!h]
\includegraphics[width=0.98\columnwidth]{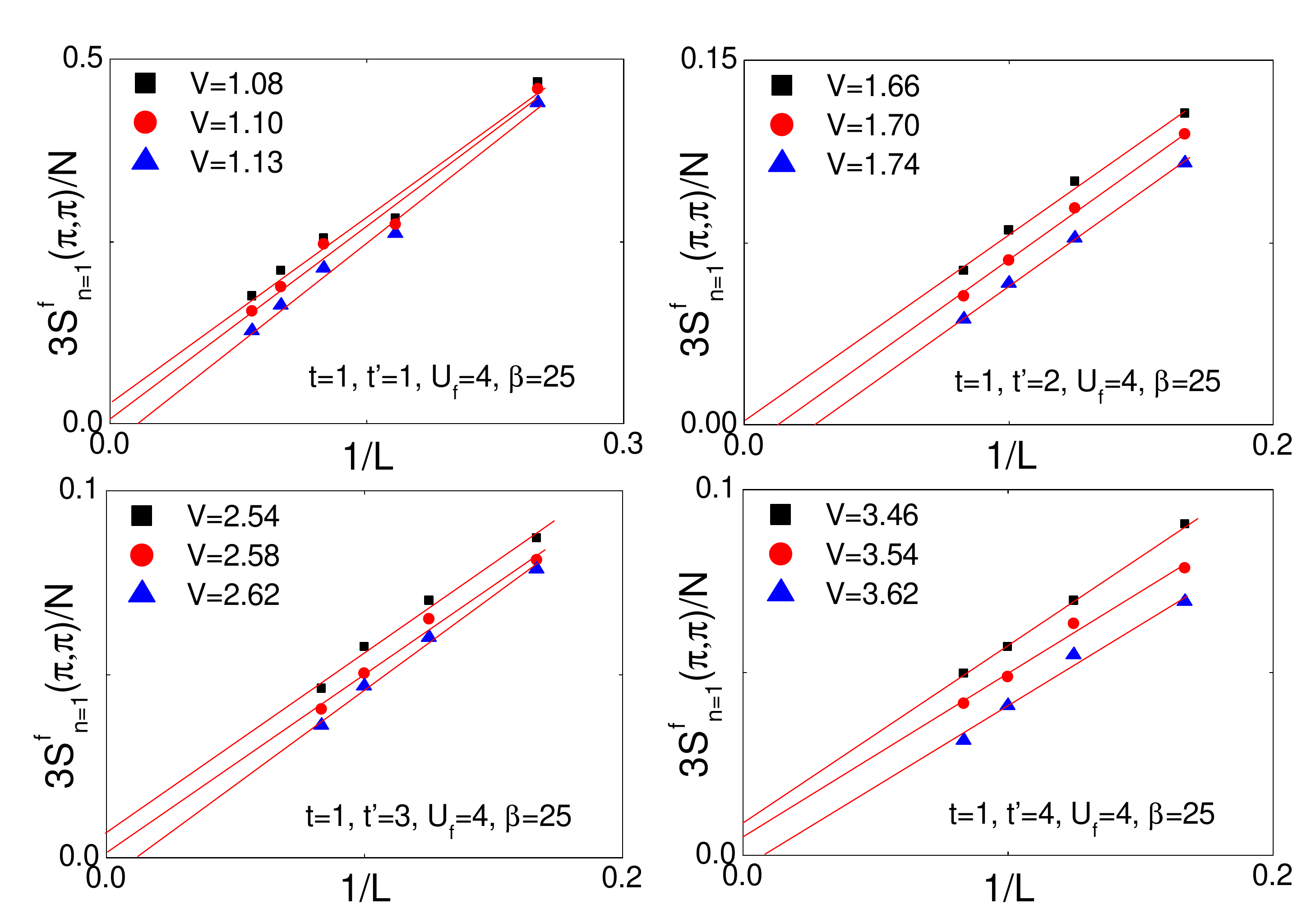} 
\caption{
Finite size scaling of the structure factor $S^f_{n=1}(\textbf{Q})$ at
$t'=1$, $2$, $3$, $4$ and fixed $U_f=4, \beta=25$. 
The subscript $n=1$ emphasizes we 
exclude $C^f\big({\bf r}=(0,0)\big)$ to reduce the
finite size corrections.  See text.  
The critical $V$ above which AF order is lost (the extrapolation becomes
zero) increases with $t'$.
The extrapolation was performed
using a linear least-squares fit.
\label{fig:KIM_scaling}
}
\end{figure}

If there is long range AF order in the system, $C^f(\textbf{r})$ remains
non-zero to large separations ${\bf r}$ and hence the spatial sum to
form $S^f(\textbf{Q})$ yields a quantity which increases linearly with
the system size $N$.  Spin wave theory\cite{Huse88} provides the
analytic form for the finite size correction
\begin{align}
   \frac{3S^f(\textbf{Q})}{N} = \frac{a}{L}+m^2_{AF,f}
\label{eq:correction}
\end{align}
Here $m_{AF,f}$ is the AF order parameter in the thermodynamic limit and
$L=\sqrt{N}$ is the linear lattice size.  The
correction factor $a$ can be reduced by excluding short range
terms $C^f\big({\bf r}=(0,0)\big)$ from the sum used to build the full 
structure factor, since spin correlations at short distances are enhanced over the square of the order parameter $m_{AF,f}$.  An improved estimator (lower
finite size effects) is therefore \cite{Varney09},
\begin{align}
  S^f_n(\textbf{Q})=\frac{N}{N-n}\sum_{\textbf{r}, |r|>l_c}e^{i\textbf{Q}\textbf{r}}C^f(\textbf{r})
\label{eq:structure_exclusion}
\end{align}
where $n$ is the number of separations ${\bf r}$ shorter than $l_c$
which are excluded from the sum.  In our DQMC measurements, we chose
$l_c=0$ and $n=1$, removing only the fully local spin-spin correlation
$C^f(0,0)$.  (This is the off-vertical scale data point in 
Fig.~\ref{fig:KIM_spsp}.)

\begin{figure}[!h]
\includegraphics[width=0.98\columnwidth]{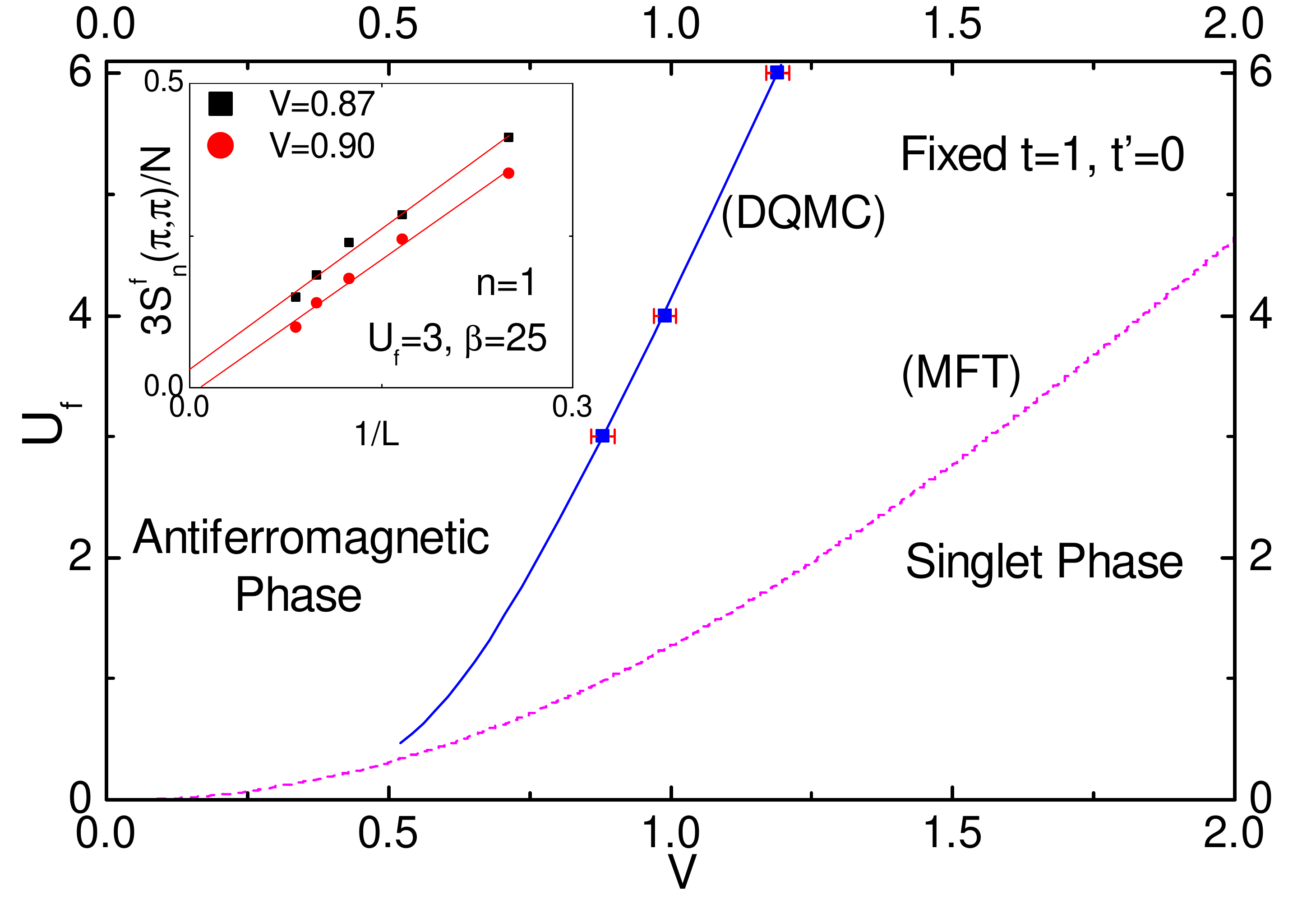} 
\caption{
The ground state $U_f$-$V$ phase diagram of the half-filled PAM
($t'=0)$.  The
three blue squares indicate the DQMC QCP, while the dotted
line shows the MFT results. The solid line is a guide to the eye.
\underline{Inset:} Finite size scaling of $S^f_n(\textbf{Q})$
for $n=1$ at
$V=0.87$ and $V=0.90$, and $U_f=3$, $\beta=25$
We have set $n=1$ corresponding to 
exclusion of  $C\big({\bf r}=(0,0)\big)$. 
The extrapolation was performed using a linear least-squares fit. 
\label{fig:PAM_phase_diagram}
}
\end{figure}

\begin{figure}[!h]
\includegraphics[width=0.98\columnwidth]{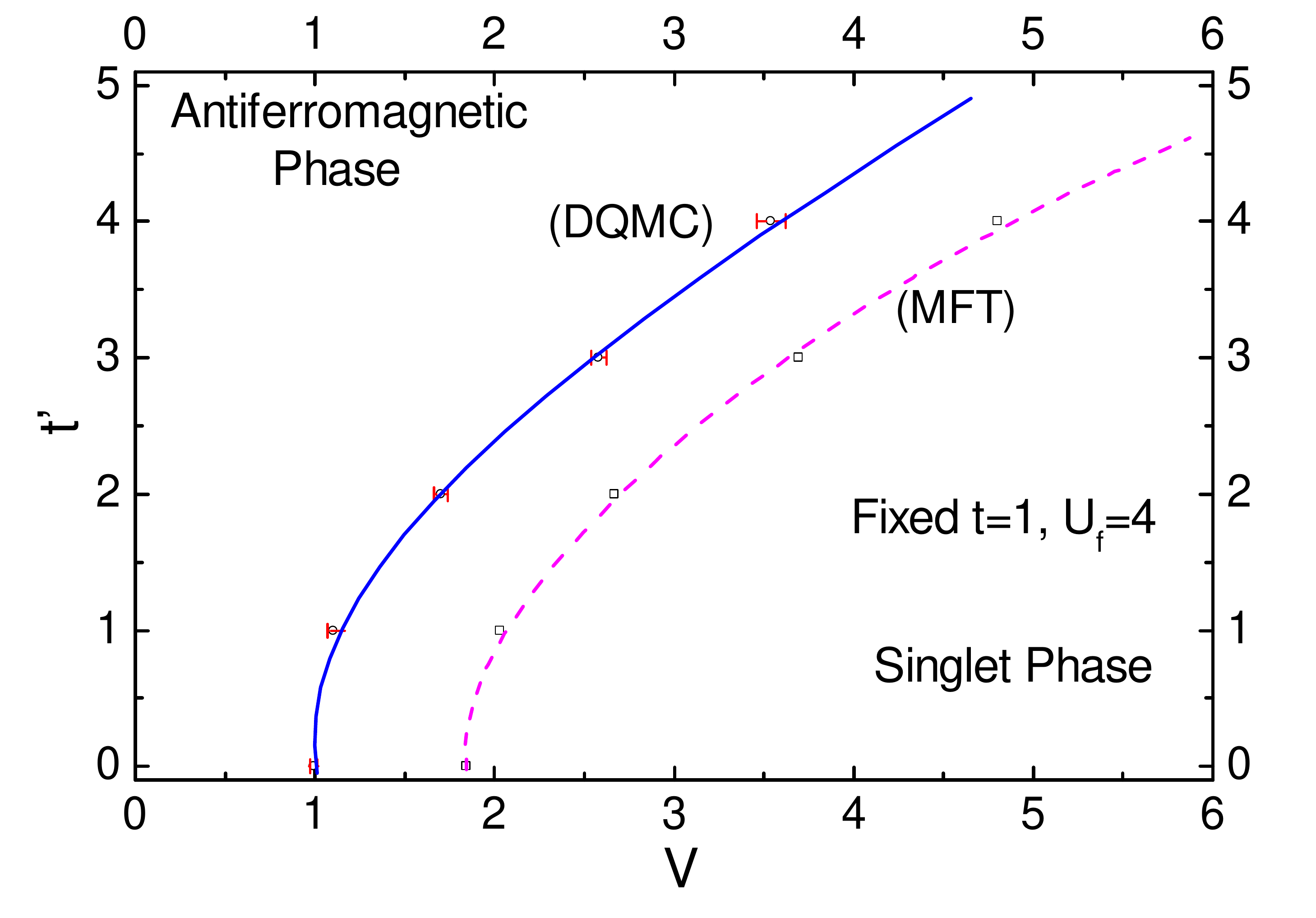} 
\caption{
The ground state $t'$-$V$ phase diagram of the half-filled KI-M system
at fixed $t=1,U_f=4$. The blue solid line indicates the DQMC boundary
and the pink dotted line indicates the MFT boundary, both of which
separate the AF phase and the singlet phase.
\label{fig:DQMC_MFT_phase_diagram}
}
\end{figure}

To locate the phase transition point accurately, we measure 
the structure factor $S^f_n(\textbf{Q})$ according to
Eq.~\ref{eq:correction}, and then extrapolate to get the order parameter
$m_{AF,f}$.  The results are shown in Fig.~\ref{fig:KIM_scaling}. Fixing
$t=1, U_f=4, \beta=25$, then for $t'=1$, 
$m_{AF,f}$ has a negative extrapolation at $V=1.13$, but is positive  at
$V=1.08$.  These bracket the QCP which we
estimate to be at  $V_c=1.10\pm 0.03$. Similarly,
for $t'=2$, we conclude $V_c=1.70 \pm 0.04$; for $t'=3$, we have
$V_c=2.58 \pm 0.04$; and finally for $t'=4$,  we find $V_c=3.54 \pm 0.08$.

Although our focus here is on the KI-M model and ascertaining the
effect of additional metallic bands on the AF-singlet transition,
we also have determined $V_c$ more accurately for the PAM ($t'=0$).
To our knowledge, the original DQMC
results\cite{Vekic95} have not been re-examined.
The main panel of
Fig.~\ref{fig:PAM_phase_diagram} shows the $U_f$-$V$ phase diagram,
and the AF and singlet regions at $t'=0$.
The critical points are deduced from the scaling of
$S^f_n(\textbf{Q})$.  A representative plot is
shown in the inset for $U_f=3$. $V=0.90$ (zero intercept) 
and $V=0.87$ (non-zero intercept), bracket a 
critical value $V_c=0.89\pm 0.02$ at $U_f=3$. Similarly, for $U_f=4$, we
conclude $V_c=0.99 \pm 0.02$ and for $U_f=6$, we conclude $V_c=1.18 \pm
0.02$.  The dotted line showing the MFT results has been discussed in
the previous section.

We now consider the phase diagram for the full KI-M Hamiltonian with $t'
\neq 0$.  Here we chose to fix $U_f=4$ and focus on the effect of
coupling the KI to the metal with $t'$.  The phase diagram is shown in
Fig.~\ref{fig:DQMC_MFT_phase_diagram}. With the increase of $t'$, the
critical value $V_c$ increases in both DQMC and MFT calculations,
quantifying the degree to which interlayer hopping parameter $t'$
enhances the RKKY interaction and stabilizes the AF phase.  The increase
in $V_c$ is quite substantial.  In contrast, previous
comparisons\cite{held00} of the PAM with on-site (insulating) and intersite (metallic) $d$ $f$ hybridization did not reveal as great a
difference in $V_c$.  This suggests the increase found here is
associated with the van-Hove singularity in the DOS.

\section{Spectral Function and Dynamic Spin Structure Factor}

DQMC is also able to evaluate (real time) dynamic information through analytic
continuation\cite{jarrell96} of the imaginary time correlation function.
Inverting the integral relation,
\begin{align}
G_\alpha(\textbf{q},\tau)=\int_{-\infty}^{+\infty}d\omega
\frac{e^{-\omega \tau}}{1+e^{-\beta \omega}}
A_\alpha(\textbf{q},\omega)
\label{aqw}
\end{align}
yields the spectral function from the one particle Greens functions measured in
DQMC.  Here $\alpha=c,d,f$ labels the band.  The associated densities of
states are given by $\rho_\alpha(\omega) = \sum_{\textbf{q}} A_\alpha(\textbf{q},\omega)$.
The low frequency behavior of $\rho_\alpha(\omega)$ quantifies the
possible existence of Slater, Mott, or hybridization gaps.

The dynamic spin structure factor is similarly related to an imaginary
time counterpart which is a generalization of the quantity of
Eq.~\ref{eq:dy_moment} to include intersite correlations, 
\begin{align}
\chi_\alpha(\textbf{q},\tau)&=
\frac{1}{N} \sum_{\textbf{j,l}}
e^{i \textbf{q} (\textbf{j}-\textbf{l})}\,
\langle S^\alpha_{\textbf{j}z}(\tau)S^\alpha_{\textbf{l}z}(0) \rangle
\nonumber \\
\chi_\alpha(\textbf{q},\tau)&=\frac{1}{\pi}\int_{-\infty}^{+\infty} d\omega
\frac{e^{-\omega \tau}}{1-e^{-\beta \omega}} \textup{Im} \chi_\alpha(\textbf{q},\omega) \label{chiqw}
\end{align} 
In an AF ordered phase, the presence of low energy spin wave
excitations leads to a vanishing of the gap in $ \textup{Im} \chi_\alpha(\textbf{q},\omega)$
at the ordering wave vector (in our case ${\bf Q}=(\pi,\pi)$).

\begin{figure}[!h]
\includegraphics[width=0.98\columnwidth]{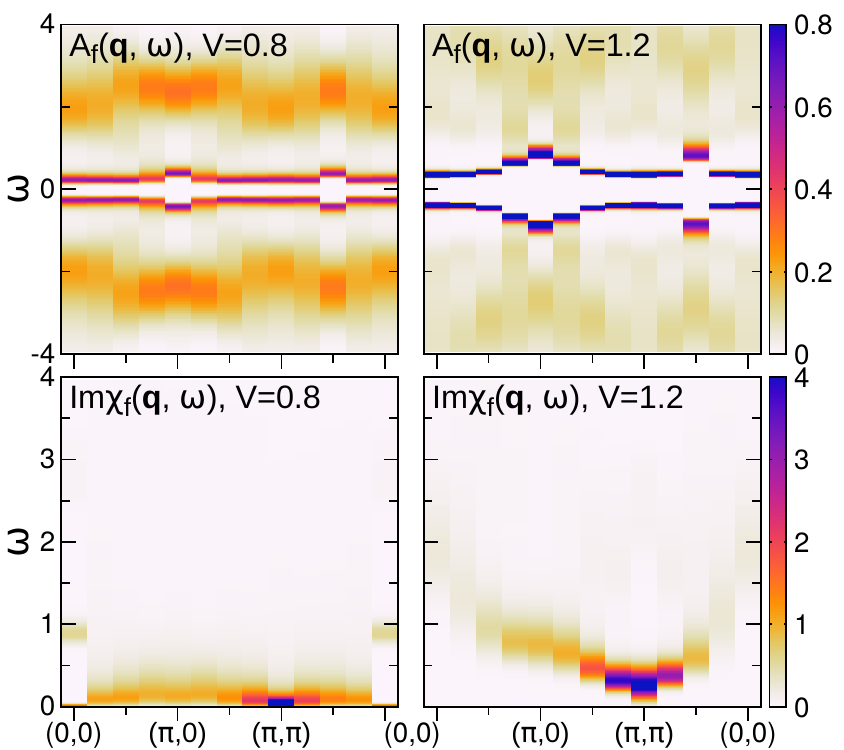} 
\caption{\underbar{Top row:} 
The one particle spectral function $A_f(\textbf{q}, \omega)$ of the PAM
in the
presence (left) and absence (right) of AF order. 
\underbar{Bottom row:} 
Spin spectral function $ \textup{Im}\chi_f(\textbf{q}, \omega)$. Results
are computed on an $N=8\times 8$ cluster at fixed $t=1,
U_f=4, \beta=25$. 
\label{fig:PAM_A_Chi}
}
\end{figure}

\begin{figure}[!h]
\includegraphics[width=0.98\columnwidth]{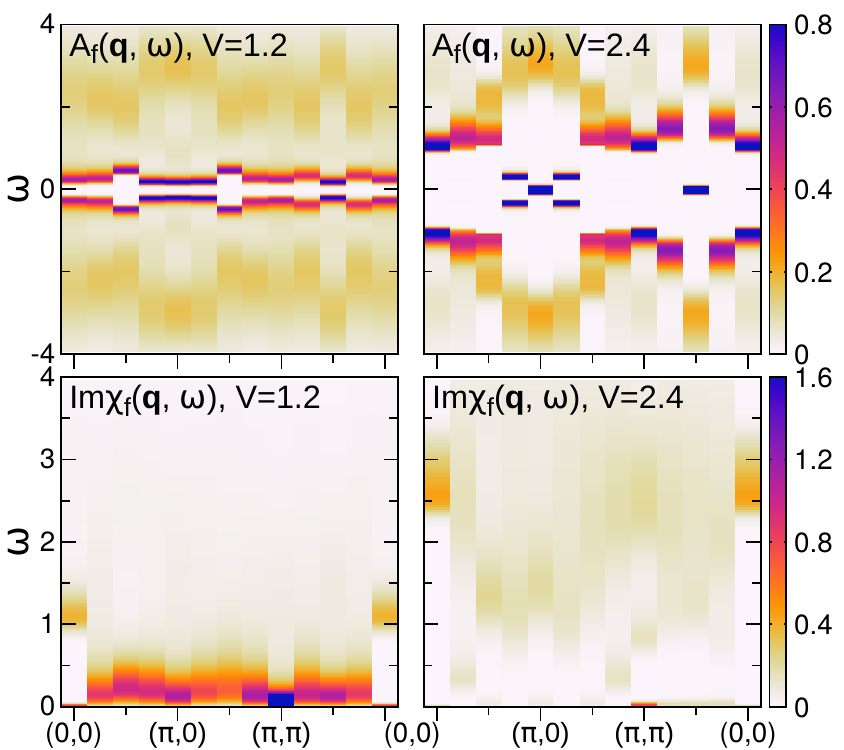} 
\caption{
\underbar{Top row:} 
The one particle spectral function $A_f(\textbf{q}, \omega)$ of the
KI-M in the
presence (left) and absence (right) of AF order. 
\underbar{Bottom row:}  
Spin spectral function $ \textup{Im}\chi_f(\textbf{q}, \omega)$. Results
are computed on an $N=8\times 8$ cluster at fixed
$t=1, t'=2, U_f=4, \beta=25$.
\label{fig:MPAM_A_Chi}
}
\end{figure}

\begin{figure}[!h]
\includegraphics[width=0.98\columnwidth]{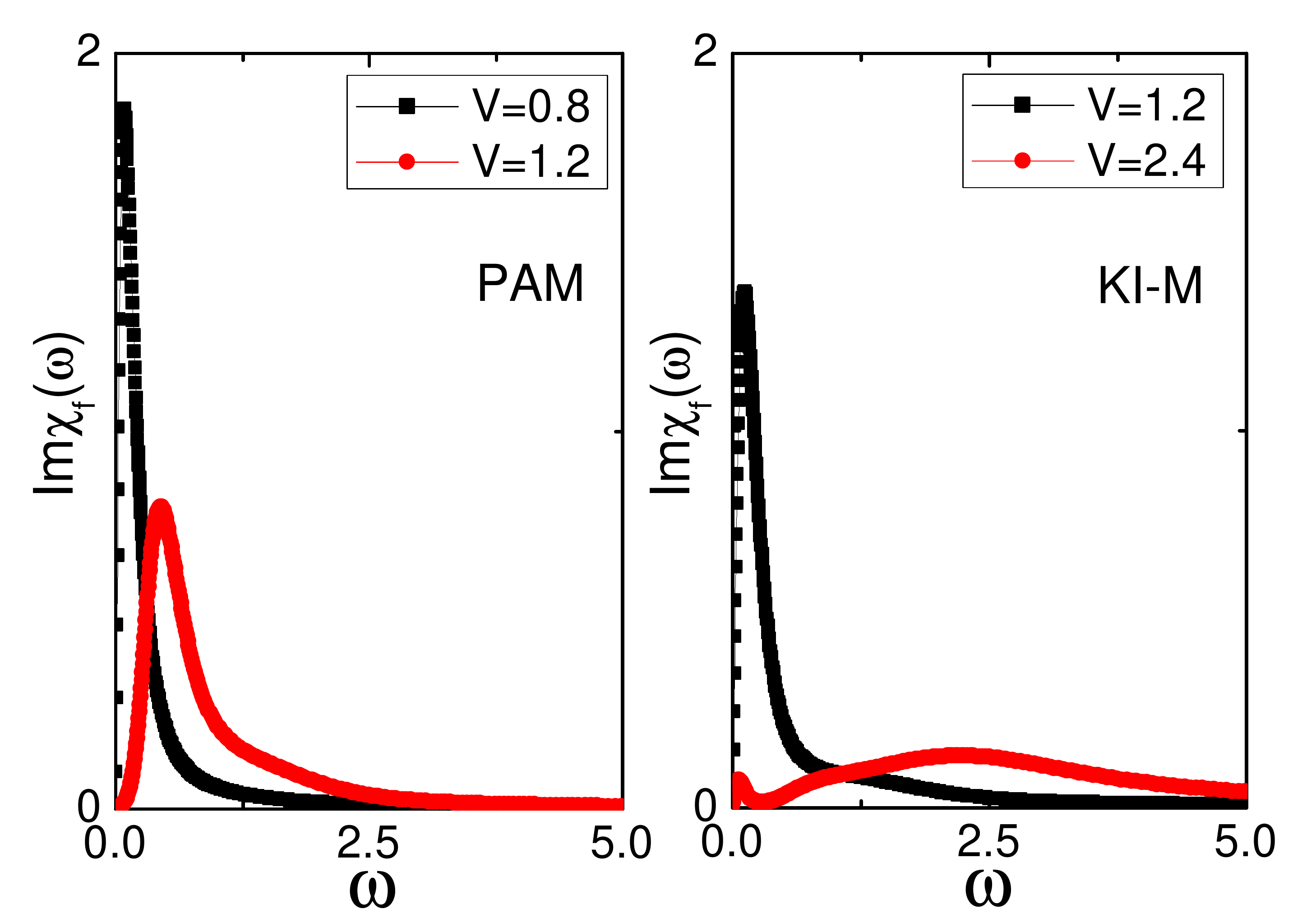} 
\caption{
\underline{Left}: The dynamic spin structure factor $ \textup{Im}\chi_f(\omega)$ of the 
PAM at fixed $t=1, U_f=4, \beta=25$.
\underline{Right}: The dynamic spin structure factor $ \textup{Im}\chi_f(\omega)$ of the KI-M at fixed $t=1, t'=2, U_f=4, \beta=25$. Both results are calculated on $N=8\times 8$ clusters.
\label{fig:MPAM_Chi}
}
\end{figure}

\vskip0.10in
In Fig.~\ref{fig:PAM_A_Chi} and
Fig.~\ref{fig:MPAM_A_Chi}, we show the one particle spectral
function $A_f(\textbf{q}, \omega)$ and the spin spectral function $ \textup{Im}\chi_f(\textbf{q},
\omega)$, which are calculated using the maximum entropy
method\citep{Gubernatis, Beach}, for the PAM model and the KI-M model
respectively. $\textup{Im}\chi_f(\textbf{q},\omega)$ complements the data
for the equal time spin and singlet correlators of Fig.~\ref{fig:KIM_spsp}.
In their AF phases (left panels) 
the PAM and the KI-M models are both characterized by
a single particle gap in $A_f(\textbf{q}, \omega)$.
$ \textup{Im}\chi_f(\textbf{q},\omega)$ has a finite spectral weight
(no gap) near 
$\omega=0$ indicating the presence of low energy spin wave excitations in
an AF phase. 
As has previously been noted in DQMC\cite{Vekic95} 
in the singlet phase of the PAM, 
a spin gap opens in $ \textup{Im}\chi_f(\textbf{q},\omega)$.
Figure \ref{fig:MPAM_Chi} gives the momentum integrated
$ \textup{Im}\chi_f(\omega)$.  The similarity between the PAM and KI-M
is clear, although the singlet phase dynamic spin response
of the KI-M is considerably broader, a natural consequence of the larger 
value of $V$ required to destroy AF order and of the
hybridization to an additional band.

The singlet phase of the KI-M is distinguished from the PAM 
by a non-zero  $A_f(\textbf{q},
\omega)$ at $\omega=0$
and also a very broadly distributed spectral weight $ \textup{Im}\chi_f(\textbf{q},\omega)$
(Fig.~\ref{fig:MPAM_A_Chi}, bottom right).
The distinction between the singlet phases of the KI-M and the PAM is further confirmed by the dynamic spin structure factors, given by $ \textup{Im}\chi_f(\omega) = \sum_{\textbf{q}}  \textup{Im}\chi_f(\textbf{q},\omega)$ for both models, as shown in Fig.~\ref{fig:MPAM_Chi}.

\vskip0.10in
\section{Conclusions}

A considerable body of existing theoretical and numerical work
\cite{potthoff95a,potthoff95b,okamoto04,zenia09,ishida12,helmes08,Euverte12}
has examined coupling of a single band Hubbard model to additional
conduction electrons as a model of metal-insulator interfaces,
and the possibility of penetration of AF and Mott insulator 
features of strong interaction into the metal, and {\it vice-versa}.
Qualitative similarities exist between phenomena like singlet
formation between electrons in distinct bands and
between electrons in two conjoined materials.
In this paper we have first shown that within a self-consistent MFT 
there is a tendency towards expansion of the region of AF stability
in a three band extension of the PAM.



We next employed the DQMC method to confirm these findings with an
exact, beyond MF, 
treatment of the correlations, and thereby identify quantitatively the
critical $f$ $d$ hybridization in the plane of interaction strength $U_f$ and
hopping $t'$ between the PAM and the metal.  In the process, we improved
on the previously known $V_c$ in the PAM ($t'=0$) limit.  Our primary
observables in the characterization of the phases were the AF structure
factor, the singlet correlator, and the dynamical moment, which all
provide a consistent picture of the location of the phase boundary.
Work within DMFT\cite{peters13}, which focuses on the paramagnetic
phase, is complementary to what we have done here.

Although the AF phase of the KI-M is stabilized by contact with the
metal, the behavior of $ \textup{Im}\chi_f(\textbf{q},\omega)$ is not dramatically
different from the PAM.  $ \textup{Im}\chi_f(\textbf{q},\omega)$ has a gap $\Delta_s$ at low
freqencies in the singlet phase, but has non-zero low 
frequency spectral weight in the AF phase associated with spin-wave
excitations.  

In contrast,
the single particle spectral weight $A_f(\textbf{q},\omega)$,
and the momentum-integrated density of states,
behave differently in the KI-M than the PAM.  The PAM has a nonzero 
charge gap $\Delta_c$ in both the
AF and singlet phases\cite{Vekic95}, with 
$\Delta_s/\Delta_c \rightarrow 1$  as $V$ increases to deep in
the singlet phase.  We find here that for the KI-M there
are peaks in $A(\textbf{q},\omega)$ near $\textbf{q}=(\pi,0)$ and $\textbf{q}=(\pi/2,\pi/2)$
and hence also in $\rho_f(\omega)=\sum_{\textbf{q}} A_f(\textbf{q},\omega)$.
We believe this distinction to originate in the fact that
even though the AF order is lost, the additional $c$-electrons
still strongly interact with the $d$ and $f$ bands of the KI, so
that there is no longer an insulating Kondo gap.

The Hamiltonian, Eq.~\ref{eq:ham} includes $cd$ and $df$ hybridizations.
We have also done some studies of the effect of on-site $cf$ hopping.
In order to keep the lattice bipartite and avoid a sign problem we have
altered the $df$ hybridization to a near-neighbor form also used, for
example, in \cite{held00}.  This change does not shift the critical $V$
from the on-site value, to within our error bars.  Having verified this,
we then added $cf$ hopping and find that it, also, leaves the critical
point at the same value.  We conclude that more complex (and realistic)
forms of the electronic kinetic energy have little effect on the
qualitative and quantitative results of our paper:  an enhancement of
the regime of AF order.

Our work on a two layer metal-PAM represents a first step in the
application of DQMC to the more general investigation of $f$-electron-metal superlattices, where, on the experimental side,
dimensionality can be controlled\cite{shishido10}.  Theoretical and
numerical studies within dynamical mean field
theory\cite{peters13,okamoto08,tada13,peters14} of these structures have
already led to great insight.  Including the full spatial structure of
each layer, as done in DQMC, makes the full superlattice problem
challenging.


\begin{acknowledgements}
We acknowledge N.C. Costa for useful discussions. WH and RTS at UC-Davis were supported by the U.S. Department of Energy under grant number DE-SC0014671.  EWH and BM at Stanford/SLAC, primarily for the analytic continuation and its interpretation, were supported by the U.S. Department of Energy (DOE), Office of Basic Energy Sciences, Division of Materials Sciences and Engineering, under Contract No. DE-AC02-76SF00515.
\end{acknowledgements}


\begin{thebibliography}{100}

\bibitem{Anderson61}
``Localized Magnetic States in Metals",
P W. Anderson, Phys. Rev. 124, 41 (1961).

\bibitem{stewart84}
``Heavy-fermion systems",
G.R. Stewart, Rev. Mod. Phys. 56, 755 (1984).

\bibitem{lee86}
``Theories of heavy-electron systems",
P.A. Lee, T.M. Rice, J.W. Serene, L.J. Sham, and J.W. Wilkins,
Comments Condens. Matt. Phys. 12, 99 (1986).

\bibitem{Georges96}
``Dynamical mean-field theory of strongly correlated fermion systems and the limit of infinite dimensions",
A. Georges, G. Kotliar, W. Krauth, and M.J. Rozenberg, 
Rev. Mod. Phys. 68, 13 (1996).
 
\bibitem{Vidhyadhiraja04}
``Dynamics and scaling in the periodic Anderson model",
N.S. Vidhyadhiraja and D E. Logan, Eur. Phys. J. B 39, 313 (2004). 

\bibitem{Sen16}
``Quantum critical Mott transitions in a bilayer Kondo insulator-metal model system",
S. Sen and N. S. Vidhyadhiraja, Phys. Rev. B 93, 155136 (2016).

\bibitem{hewson93}
A.C. Hewson, {\it The Kondo Problem to Heavy Fermions},
(Cambridge University Press, Cambridge, 1993).

\bibitem{Shinzaki16}
``DMFT Study for Valence Fluctuations in the Extended Periodic Anderson Model",
R. Shinzaki, J. Nasu, and A. Koga: J. Phys. Conf. Ser. 683, 012041 (2016).

\bibitem{allen82}
``Kondo Volume Collapse and the $\gamma\rightarrow\alpha$ Transition in Cerium",
J.W. Allen and R.M. Martin,
Phys. Rev. Lett. 49, 1106 (1982).

\bibitem{allen92}
``$\alpha \rightarrow \gamma$ transition in Ce. II. A detailed analysis
of the Kondo volume-collapse model", J.W. Allen and L.Z. Liu, Phys. Rev.
B 46, 5047 (1992).

\bibitem{lavagna83}
``Volume collapse in the Kondo lattice",
M. Lavagna, C. Lacroix, and M. Cyrot,
Phys. Lett. A 90, 210 (1982).

\bibitem{gunnarsson83}
``Electron spectroscopies for Ce compounds in the impurity model",
O. Gunnarsson and K. Schonhammer,
Phys. Rev. B 28, 4315 (1983).

\bibitem{mcmahan98}
``Volume Collapse transitions in the rare earth metals,"
A. McMahan, C.~Huscroft, R.T.~Scalettar, and E.L.~Pollock,
J. Comput.-Aided Mater. Des. 5, 131 (1998).

\bibitem{lipp08}
``Thermal Signatures of the Kondo Volume Collapse in Cerium"
M.J. Lipp, D. Jackson, H. Cynn, C. Aracne, W.J. Evans, and A.K. McMahan,
Phys. Rev. Lett. 101, 165703 (2008).

\bibitem{bradley12}
``4f electron delocalization and volume collapse in praseodymium
metal,"
J.A. Bradley, K.T. Moore, M.J. Lipp, B.A. Mattern,
J.I. Pacold, G.T. Seidler, P. Chow, E. Rod, Y. Xiao,
and W.J. Evans
Phys. Rev. B 85, 100102(R) (2012).

\bibitem{lanata15}
``Phase Diagram and Electronic Structure of Praseodymium and Plutonium,"
N. Lanat\'a, Y. Yao, C-Z. Wang, K-M. Ho, and G. Kotliar,
Phys. Rev. X 5, 011008 (2015).

\bibitem{pixley15}
``Pairing correlations near a Kondo-destruction quantum critical
point,"
J.H. Pixley, L. Deng, K. Ingersent, and Q. Si,
Phys. Rev. B 91, 201109(R) (2015).

\bibitem{Ruderman54}
``Indirect Exchange Coupling of Nuclear Magnetic Moments by Conduction Electrons",
M.A. Ruderman and C. Kittel, Phys. Rev. 96, 99 (1954).

\bibitem{Kasuya56}
``A Theory of Metallic Ferro- and Antiferromagnetism on Zener's Model",
T. Kasuya, Prog. Theor. Phys. 16, 45 (1956).

\bibitem{Yosida57}
``Magnetic Properties of Cu-Mn Alloys",
K. Yosida, Phys. Rev. 106, 893 (1957).

\bibitem{Kondo69}
``Theory of Dilute Magnetic Alloys",
J. Kondo, Solid State Phys. 23, 183 (1969).

\bibitem{Kouwenhoven01}
``Revival of the Kondo effect",
L. Kouwenhoven and L. Glazman, Phys. World 14, No. 1, 33-38 (2001).

\bibitem{mannhart05}
J. Mannhart and D.G. Schlom,
in {\it Thin Films and Heterostructures for
Oxide Electronics}, edited by S. ogale
(Springer, New York, 2005), p25.

\bibitem{mannhart10}
``Oxide Interfaces-  An Opportunity for Electronics",
J. Mannhart and D.G. Schlom,
Science 327, 1607 (2010).

\bibitem{millis05}
A. Millis, in {\it Thin Films and Heterostructures for
Oxide Electronics}, edited by S. Ogale
(Springer, New York, 2005), p279.

\bibitem{freericks06}
J.K. Freericks, {\it Transport in Multilayers Nanostructures:
The Dynamical Mean Field Theory Approach},
(Imperial College Press, London, 2006).

\bibitem{Euverte12}
``Kondo Screening and Magnetism at Interfaces,"
A. Euverte, F. H\'ebert, S. Chiesa, R.T. Scalettar, 
and G. G. Batrouni, Phys. Rev. Lett. 108, 246401 (2012)

\bibitem{shishido10}
``Tuning the Dimensionality of the Heavy Fermion Compound CeIn$_3$",
H. Shishido, T. Shibauchi, K. Yasu, T. Kato, H. Kontani,
T. Terashima, and Y. Matsuda,
Science 327, 980 (2010).


\bibitem{peters13}
``Kondo effect in $f$-electron superlattices,"
R. Peters, Y. Tada, and N. Kawakami,
Phys. Rev. B 88, 155134 (2013).

\bibitem{liebsch04}
``Single Mott transition in the multiorbital Hubbard model",
A. Liebsch, Phys. Rev. B 70, 165103 (2004).

\bibitem{arita05}
``Orbital-selective Mott-Hubbard transition in the two-band Hubbard model",
R. Arita and K. Held, Phys. Rev. B 72, 201102(R) (2005).

\bibitem{Zhang13}
``Periodic Anderson model with electron-phonon correlated conduction band",
P. Zhang, P. Reis, K. Tam, M. Jarrell, J. Moreno, 
F. Assaad, and A. K. McMahan, Phys. Rev. B 87, 121102(R) (2013).

\bibitem{Sen15}
``Spectral changes in layered f-electron systems induced by Kondo hole substitution in the boundary layer",
S. Sen, J. Moreno, M. Jarrell, and N.S. Vidhyadhiraja, 
Phys. Rev. B 91, 155146 (2015).

\bibitem{Euverte13}
``Finite f-electron Bandwidth in a Heavy Fermion Model",
A. Euverte, S. Chiesa, R.T. Scalettar, and G.G. Batrouni,
Phys. Rev. B 88, 235123 (2013).

\bibitem{Blankenbecler81}
``Monte Carlo calculations of coupled boson-fermion systems. I",
R. Blankenbecler, D.J. Scalapino, and R.L. Sugar, 
Phys. Rev. D 24, 2278 (1981).

\bibitem{Loh92}
E. Loh and J. Gubernatis, in Modern Problems of Condensed Matter 
Physics, edited by W. Hanke and Y. V. Kopaev (North Holland, 
Amsterdam, 1992), Vol. 32, p. 177.

\bibitem{Vekic95}
``Competition Between Antiferromagnetic Order and Spin Liquid Behavior
in the Two--Dimensional Periodic Anderson Model at Half--Filling,"
M. Vekic, J.W. Cannon, D.J. Scalapino, R.T. Scalettar, 
and R.L. Sugar, Phys. Rev. Lett. 74, 2367 (1995).

\bibitem{bernhard15}
``Coexistence of magnetic order and Kondo effect in the Kondo-Heisenberg
model",
B.H. Bernhard and C. Lacroix,
Phys. Rev. B92, 094401 (2015).

\bibitem{eder16}
``Antiferromagnetic phases of the Kondo lattice,"
R. Eder, K. Grube, and P. Wr\'obel,
Phys. Rev. B93, 165111 (2016).


\bibitem{Costaa}
``Spiral magnetic phases on the Kondo Lattice Model: A Hartree–Fock approach",
N.C. Costaa, J.P. Limab and R. R. dos Santosa, Journal of Magnetism and Magnetic Materials 423 (2017).

\bibitem{Huscroft99}
``Magnetic and Thermodynamic Properties of the Three-Dimensional
periodic Anderson Hamiltonian,"
C. Huscroft, A.K. McMahan, and R.T. Scalettar, Phys. Rev. Lett. 82, 
2342 (1999).

\bibitem{Huse88}
``Ground-state staggered magnetization of two-dimensional quantum Heisenberg antiferromagnets",
D.A. Huse, Phys. Rev. B 37, 2380 (1988). 

\bibitem{Varney09}
``Quantum Monte Carlo Study of the 2D Fermion Hubbard
Model at Half-Filling,"
C.N. Varney, C.-R. Lee, Z.J. Bai, S. Chiesa, M. Jarrell, and 
R.T. Scalettar, Phys. Rev. B 80, 075116 (2009).

\bibitem{held00}
``Similarities between  the Hubbard and Periodic
Anderson Models at  Finite Temperatures",
K. Held, C. Huscroft, R.T. Scalettar, and A.K. McMahan,
Phys. Rev. Lett. {\bf 85}, 373 (2000).

\bibitem{jarrell96}
``Bayesian inference and the analytic continuation of imaginary-time quantum Monte Carlo data",
M. Jarrell and J.E. Gubernatis,
Phys. Rep. 269, 135 (1996).

\bibitem{Gubernatis}
``Quantum Monte Carlo simulations and maximum entropy: Dynamics from imaginary-time data",
J. E. Gubernatis, M. Jarrell, R. N. Silver, and D. S. Sivia, Phys. Rev. B 44, 6011 (1991).

\bibitem{Beach}
``Identifying the maximum entropy method as a special limit of stochastic analytic continuation",
K. S. D. Beach, arXiv:cond-mat/0403055.

\bibitem{ohmoto02}
``Artificial Charge-Modulation in Atomic-Scale Perovskite Titanate Superlattices",
A. Ohmoto, D. Muller, J. Grazul, and H. Hwang,
Nature (London) 419, 378 (2002).

\bibitem{potthoff95a}
``Metallic surface of a Mott insulator–Mott insulating surface of a metal",
M. Potthoff and W. Nolting,
Phys. Rev. B 60, 7834 (1999).

\bibitem{potthoff95b}
``Surface metal-insulator transition in the Hubbard model",
M. Potthoff and W. Nolting,
Phys. Rev. B 59, 2549 (1999).

\bibitem{okamoto04}
``Spatial inhomogeneity and strong correlation physics: A dynamical mean-field study of a model Mott-insulator-band-insulator heterostructure",
S. Okamoto and A.J. Millis,
Phys. Rev. B 70, 241104 (2004).

\bibitem{zenia09}
``Charge-Orbital Density Wave and Superconductivity in the Strong Spin-Orbit Coupled IrTe2∶Pd",
H. Zenia, J.K. Freericks, H.R. Krishnamurthy, and
T. Pruschke,
Phys. Rev. Lett. 103, 116402 (2009).

\bibitem{ishida12}
``First-order metal-to-metal phase transition and non-Fermi-liquid
behavior in a two-dimensional Mott insulating layer adsorbed on a metal
substrate",
H. Ishida and A. Liebsch, 
Phys. Rev. B 85, 045112 (2012).

\bibitem{helmes08}
``Kondo Proximity Effect: How Does a Metal Penetrate into a Mott Insulator?",
R.W. Helmes, T.A. Costi, and A. Rosch,
Phys. Rev. Lett. 101, 066802 (2008).

\bibitem{okamoto08}
``Enhanced Superconductivity in Superlattices of High-Tc Cuprates",
S. Okamoto and T.A. Maier,
Phys. Rev. Lett. 101, 156401 (2008).

\bibitem{tada13}
``Dimensional crossover in layered f-electron superlattices",
Y. Tada, R. Peters, and M. Oshikawa,
Phys. Rev. B 88, 235121 (2013).

\bibitem{peters14}
``Surface density of states of layered f-electron materials",
R. Peters and N. Kawakami,
Phys. Rev. B 89, 041106 (2014).

\end{thebibliography}
\end{document}